\documentstyle[11pt,paspconf]{article}

\begin{document}

\title{The Core of the Great Attractor}

\author{Patrick A.\ Woudt}
\affil{European Southern Observatory, Karl-Schwarzschild-strasse 2, D-85748,
Garching bei M\"unchen, Germany}

\author{Ren\'ee C.\ Kraan-Korteweg}
\affil{Departamento de Astronom{\'{\i}}a, Universidad de Guanajuato, Apartado 
Postal 144, Guanajuato, Gto 36000, Mexico}

\author{Anthony P.\ Fairall}
\affil{Department of Astronomy, University of Cape Town,
Rondebosch 7700, South Africa}

\begin{abstract}
The nature and extent of the Great Attractor (GA) has been the subject
of much debate in the past decade, partly due to the fact that a large
fraction of the GA overdensity is hidden by the southern Milky Way.
Based on our deep optical galaxy search behind the southern Milky Way
and a subsequent redshift survey we discovered that the Norma cluster
(ACO 3627) in the GA region is a very massive cluster of galaxies.
The cluster is comparable in size, richness and mass to the Coma
cluster. It is located at the intersection of two distinct large
structures, the Centaurus Wall and the Norma Supercluster. 
The velocity flow fields in the GA region are most
likely caused by the confluence of these two massive structures
where the Norma cluster constitutes its previously unseen but
predicted core. The possibility that another, heavily obscured and yet
uncharted rich cluster might form part of the GA overdensity is also
discussed.
\end{abstract}


\section{The Zone of Avoidance; the Milky Way as a natural barrier}

Optical galaxy catalogues become severely incomplete towards the
Galactic Equator due to the absorbing dust in the plane of the Milky
Way which increasingly reduces the magnitudes and isophotal diameters
of external galaxies. In the optical, as much as 20\% of the
extragalactic sky is obscured by the Galaxy.

This incompleteness severely constrains the studies of large-scale
structures in the nearby Universe, the gravitational acceleration on
the Local Group and other streaming motions, in particular in the
region of the Great Attractor (GA). This nearby mass overdensity
at ($\ell, b, v$) $\sim$ ($320{\deg}, 0{\deg}, 4500$ km/s) (Kolatt et
al.\ 1995) -- centered in the Galactic Plane! -- is largely hidden by
the southern Milky Way.

The ambiguity about the nature and extent of the GA arises from the
less than perfect match between the reconstructed mass density field
and the galaxy density field (Dekel 1994). In other words, the peak in
the mass density field is not always reflected by the galaxy
distribution in redshift space. This raises the question whether light
traces mass, or whether something else induces the observed
streaming motions. The very existence of the GA has even been
questioned (Rowan-Robinson 1993). Observational evidence for a Great
Attractor, however, has grown (eg.\ the SBF Survey, Tonry et al., this
volume).

Observing the galaxy distribution in the ZOA through multi-wavelength 
studies -- optical, near-infrared, HI, X-ray -- will in due time 
provide a clear answer on the various components of the GA. Is 
there still a considerable mass concentration hidden by the Milky Way 
(as suggested by the POTENT analyses)? Or is the GA associated with the 
Centaurus--Hydra--Pavo supercluster solely, without an additional massive 
component in the ZOA (Rowan-Robinson et al.\ 1990)? 

\section{Lifting the obscuring veil of the southern Milky Way}

To address these questions, we have performed a deep optical search
for partially obscured but still visible galaxies behind the Milky Way
(Kraan-Korteweg 1989, Kraan-Korteweg and Woudt 1994a, Woudt 1998).
Inspecting the IIIaJ film copies of the SRC Sky Survey, we identified
over 8000 galaxies with D $\ge$ 0.2 arcminutes in the general
direction of the GA overdensity ($285\deg \le \ell \le 340\deg$ and
$-10\deg \le b \le 10\deg$, Woudt 1998), the majority of which were
previously unknown (97\%).

This galaxy search significantly reduces the Zone of Avoidance:
analyzing our diameter completeness limit as a function of the Galactic
foreground extinction -- as indicated by the DIRBE/IRAS reddening maps 
(Schlegel et al.\ 1998) -- we could show that our galaxy catalogues
are complete for galaxies with {\sl extinction-corrected} diameters 
${\rm D^o} \ge 1\farcm3$ down to extinction levels in the blue of 
A$_{\rm B} \le 3{\fm}0$, respectively Galactic latitudes 
$|\, b\, | \ga 3\deg-4\deg$ of the area surveyed here. 

The main outcome of this galaxy search in the GA region
is the recognition that the Norma cluster (ACO 3627, Abell et al.\
1989) in the Zone of Avoidance (a) is a rich and very massive cluster of 
galaxies, (b) is located at the core of the Great Attractor 
(Kraan-Korteweg et al.\ 1996, Woudt 1998), and (c) is comparable 
in mass and size to the well-known Coma cluster but even nearer in
redshift space ($\ell, b, v$) = ($325.3{\deg}, -7.2{\deg}, 4844$ km/s).

\section{The Norma cluster (ACO 3627)}

In this section, we describe the main properties of this
cluster. Within the Abell radius of the Norma cluster (3 h$_{50}^{-1}$
Mpc), 603 galaxies with a major diameter ${\rm D} \ge 0{\farcm}2$ were
discovered in our galaxy search. Within the core radius, a large fraction
(48\%) are early type galaxies (Woudt 1998).

With our follow-up redshift survey, redshifts were obtained for 266
(44.1\%) of the galaxies in the Abell radius (only 17 redshifts were
known before our survey).  219 are likely cluster members. The
velocity distribution of these galaxies is close to Gaussian. A more
detailed statistical analysis (following Pinkney et al.\ 1996)
reveals, however, significant substructure within the Abell radius:
the Norma cluster consists of a main cluster at a redshift of 4844
$\pm$ 63 km/s with a velocity dispersion of 848 km/s, and of a
spiral-rich subcluster falling into the main cluster (Woudt 1998).

With these new observations, we can now clearly identify two distinct
structures that cross the Galactic Plane in the Great Attractor
region:

$\bullet$ The Centaurus Wall (Fairall 1998), tilted about 15 degrees to the
Supergalactic Plane. It extends in redshift-space from 0 - 6000 km/s and
includes the Virgo Supercluster, the Centaurus cluster and the Norma cluster.


$\bullet$ A massive broad structure between $4000 \le v \le 8000$
km/s.  This broad structure, dubbed the Norma Supercluster, runs at a
slight angle with respect to the Galactic Plane. It can be traced from
the Pavo cluster (Fairall et al.\ 1998) to the Norma cluster (its
centre) where -- at slightly higher redshifts -- it bends towards 
the Vela Supercluster (Kraan-Korteweg and Woudt 1994b) .

The observed peculiar velocity field in the Great Attractor most
likely results from the confluence of these two massive
structures. The Norma cluster is located at the intersection of these
two structures.  With a mass of $0.9 \times 10^{15}$ M$_{\odot}$
(within a 3 h$_{50}^{-1}$ Mpc radius) it is the most massive cluster
of galaxies in the Great Attractor overdensity -- and on par with the
well-known Coma cluster.  Simulations show that the Coma cluster would
appear practically identical to ACO 3627 if it were located at the
same position, ie.\ at the same redshift distance and subjected to the
same mean foreground extinction (A$_{\rm B} = 1\fm{1}$, Woudt 1998).

An R$_{\rm C}$ and I$_{\rm C}$ Tully-Fisher analysis of the Norma
cluster yields a relative distance modulus to the Virgo cluster of
$(m-M)_{\rm Norma} - (m-M)_{\rm Virgo}$ = 2$\fm$9 $\pm$ 0$\fm$2 and a
peculiar motion of the cluster with respect to the rest frame of the
Cosmic Microwave Background radiation of 461 $\pm$ 410 km/s (Woudt
1998).  Our data do not confirm the previously reported peculiar
motion of ACO 3627 of 1760 $\pm$ 355 km/s (Mould et al.\ 1991).
Within the uncertainty of the assumed extinction correction our data
are consistent with the Norma cluster being at rest with respect to
the Cosmic Microwave Background radiation. It therefore is {\sl the}
most likely candidate for the Great Attractor's previously unseen
core.

\section{Clustering in the GA region}

One cannot exclude the possibility that other rich clusters (like the
Norma cluster) reside in the GA region, hidden behind the deepest
extinction layers of the Milky Way.  A rich cluster such as the Norma
cluster (at the distance of the Great Attractor) would be fully
obscured if the extinction at optical wavelengths is larger than
A$_{\rm B} \ga 10$~mag. This corresponds to the inner $\pm$2--3
degrees of the plane of the Milky Way.  If such a cluster does exist,
it could be detected by its X-ray emission, but also by the radio
continuum emission of a central radio source, such as PKS1610-608 in
the Norma cluster.

A suspect for such a cluster candidate is actually given with the
strong extragalactic radio source PKS1343-601 (McAdam 1991): hidden
behind 12 magnitudes of extinction (A$_{\rm B}$), this radio source
at ($\ell, b, v$) = ($309.7{\deg}, 1.7{\deg}, 3872$ km/s) could mark
the bottom of the potential well of another cluster in the GA
overdensity (Woudt 1998).  There are clear indications from the
redshift-distribution of HI-detected galaxies (Kraan-Korteweg and
Juraszek 1999) that this is indeed the case.  We furthermore have
obtained deep $I$-band images of this region with the MPG-ESO 2.2-m
telescope and the Wide Field Imager. A first inspection of the
central field of this suspected cluster revealed a large number of
galaxies -- despite the severe extinction of A$_{\rm I} = 4\fm{5}$ in
the $I$-band at the position of PKS1343-601.

\section{Conclusions}

A clearer picture of the optical galaxy distribution in the general GA
region has emerged with the recent studies of the galaxy distribution
behind the southern Milky Way; two distinct extended structures at the
redshift distance of the GA cross the Galactic Plane, ie. the
Centaurus Wall and the Norma supercluster. The Norma cluster -- found
to be a massive and rich cluster behind the southern Milky Way -- is
located at the intersection of these two massive structures and marks
the core of the GA overdensity.

\acknowledgments

We kindly acknowledge C. Balkowski, V. Cayatte and P.A. Henning for their 
contribution in the redshift follow-up observations of our deep optical
galaxy search.

\end{document}